# Confidentiality and linked data

**December 2018**

Professor Felix Ritchie
University of the West of England

Professor Jim Smith
University of the West of England

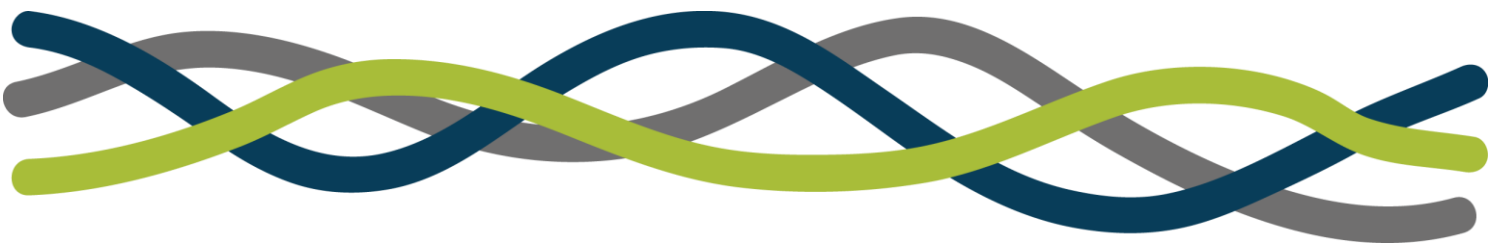

**About National Statistician's Quality Reviews (NSQR)**

National Statistician's Quality Reviews (NSQRs) cover thematic topics of national importance, conducted on behalf of and for the Government Statistical Service (GSS) in the United Kingdom. These reviews are future facing, ensuring that methods used by the GSS are keeping pace with changing data sources and technologies.

Articles contributed by leading experts published as part of the Privacy and Data Confidentiality NSQR contribute to the continuous improvement of these methods and support the GSS identify what good practice looks like for these methodologies as well as help identify opportunities for further development and investment. More information can be found on the Government Statistical Service website at: https://gss.civilservice.gov.uk/guidances/quality/nsqr/privacy-and-data-confidentiality-methods-a-national-statisticians-quality-review/. For further enquiries please contact Gentiana D. Roarson at gentiana.roarson@ons.gov.uk or the GSS Quality Centre at qualitycentre@statistics.gov.uk.

**Authors**

Professor Felix Ritchie, Bristol Centre for Economics and Finance, University of the West of England, Bristol, BS16 1QY, UK. Email: Felix.Ritchie@uwe.ac.uk

Professor Jim Smith, Computer Science Research Centre, University of the West of England, Bristol, BS16 1QY, UK. Email: James.Smith@uwe.ac.uk

**Contents**

## 1. Introduction

While survey and census data have been linked at the record level for many years, this has had a small impact on policy and research. This is partly because surveys may not contain high-quality linking information, so matching is probabilistic rather than exact, and because sampling frames may not overlap sufficiently to allow numbers of observations.

The emergence of identified administrative datasets with the potential for sharing (and thus linking) changes this significantly. The use of administrative data for policy and research offers great possibilities. The data often cover the population of interest, for example all those who have engaged with a service in some way. The data are used for delivering the service, and so are likely to be checked for accuracy. Data can be extracted from existing systems at relatively low marginal cost. The data are usually identified at source as this information is required for operations; as a result, exact matching (such as through tax or health service identifiers (IDs)) becomes more feasible and probabilistic matching more accurate. The transformation of public health analysis in recent years is a testament to the power of linking administrative data sources.

However, this potential advantage does bring additional risks for the organisation creating a linked dataset from confidential data. Freely-given informed consent is rarely given for sharing administrative data, and so the data holder may be held to a stricter standard of confidentiality than a survey data collector. The inclusion of all service users in the data reduces the level of uncertainty about whether an individual[1] is in the dataset or not. Data are likely to be matched on relatively guessable information (name, date of birth etc.). Most importantly, exact copies of (some of) the data will be held by other organisations, allowing data confidentiality to be attacked through a third party. More subtly, the third party may not regard its own data as confidential and so may leave it relatively unprotected. Finally, changes to administrative data over time can be a source of disclosure risk as users interact with services.

On the other hand, the picture may not be so bleak. Data linkage experts are more likely to identify the difficulties of effective linking as the key problem, rather than the risks of a perfect match. Data quality may be low with administrative data: cleaning is done for operational (i.e. human) rather than statistical purposes. Furthermore, administrative datasets are continuously updated, so re-identification may be restricted to a specific point in time.

This article explores these issues in more detail. The next section briefly introduces the principles and methods of linking data across different sources and points in time, focusing on potential areas of risk. We then consider confidentiality risk, focusing in particular on the 'intruder' problem central to the area, and looking at both risks from data producer outputs and from the release of microdata for further analysis. Finally, we briefly consider potential solutions to microdata release, both the statistical solutions considered in other contributed articles and non-statistical solutions.

---

[1] This discussion is couched in terms of individuals, but applies equally to organisations in, for example, business surveys. Paradoxically, the mooted replacement of data collected from surveys and imputation with data from administrative records (VAT returns etc.) may make these outputs more accurate and timely, but will decrease the protection afforded by not knowing whether a reported value is 'real' or imputed.

## 2. Methods for linking data

### 2.1 Preliminaries

In its simplest form, record linkage consists of finding matched pairs of records from two different datasets which contain different fields or identifiers, in order to create a richer picture of individuals, or to fill in missing data. A range of automated pre-processing transformations can be used to account for discrepancies or similarities in other fields (e.g. matching versions of names: 'Pete'<=>Peter <=> P). Originally reliant on hand-crafted rules, these technologies have been driven by developments in the semantic web, are both sophisticated and widely available, and are used in areas such as sales targeting.

Given two datasets $D_1$ and $D_2$, the task is therefore to divide the cross-product space $S = D_1 \times D_2$ (i.e. all possible pairs of one record from $D_1$ and one from $D_2$) into two non-overlapping areas: $M$ (containing all the matched pairs) and $U$ (containing the unmatched pairs).

- Given the potentially huge size of the search space ($|D_1| \times |D_2|$), it is extremely common to apply '**blocking'** techniques to rule out pairs that are extremely unlikely to be matched. This is effectively a pre-processing phase of assigning the label $U$ to large regions of $S$. It can be extremely effective when domain-specific knowledge can be incorporated, although it is not without risk of missing potential matches. For example, using a blocking factor to rule out pairs of individuals based on differences in 'family name' in datasets collected at different times could miss records where someone has taken their spouse's name on marriage.
- From a mathematical/machine learning perspective, linkage is done by applying a model that takes as input two records $r_i$ and $r_j$ and outputs either a label (match/unmatch) or a numerical value. In general, the two datasets may contain different information, with only a small set of common features used for matching. If $K$ denotes the set of those common features (with size K), we can define records by their values for the relevant features – e.g., $r_i = \{r_{i1}, r_{i2}, …, r_{iK}\}$.
- Rather than defining models that deal explicitly with pairs, the problem is commonly stated in terms of **comparison vectors.** Each vector has K components, denoting the differences between two records for each common feature. Since features may have different types, such as variable length text (names), dates, numbers, or specific mixes (e.g. postcodes, National Insurance (NI) numbers), we might use different measures for different features. In some cases, for example binary values, the choice of distance measure for comparing two values for a feature K may be obvious. For others, such as text fields, a number of alternatives such as different edit measures exist.
- In this formulation of the problem, deciding whether two records match consists of taking a comparison vector $C_{ij} = \{D_1(ij), D_2(ij),\ldots D_K(ij)\}$ and applying a model to deduce or predict the match value. In what follows, we will use 0 to denote that the pair of records are not matched and 1 to denote a match.
- Different approaches to record linkage are distinguished by the way that the model is specified or learned.

## 2.2 Deterministic Linkage

When those datasets each contain a common field **k\*** for an identifier that is unique to each individual, then we say that the matches found are ***deterministic***. In this case the model is human specified and is simply:

$$\text{Match }(ij) = \begin{cases} 1 & \text{if } D_{k^*}(ij) = 0, \\ 0 & \text{otherwise.} \end{cases} \qquad (1)$$

This situation applies by design in relational databases, and also by design in many internet-based records: the concept of the Semantic Web is built on the 2001 World Wide Web Consortium (W3C) definition of a Uniform Resource Identifier (URI). URL's and email addresses are examples of URIs. Calculation of matches may be time consuming if datasets are large, but this issue has been extensively tackled by computer science researchers as it forms the basis of relational database 'join' operations. If necessary, the identifiers may be encrypted by an appropriate one-way hashing algorithm: as long as both dataset owners use the same algorithm, then records with identical hashed identifiers can be matched. A variety of "Privacy Preserving Record Linkage" schemes have also been proposed that rely on the presence of a trusted third party to mediate between the different owners of datasets who are considered to be "honest but curious" [1].

## 2.3 Rule-Based Linkage

More sophisticated deterministic record linkage strategies work in an iterative fashion, for example the US National Cancer Institute use a two-stage matching strategy [2]. The first stage identifies matches based on *Social Security Number* (*SSN*) and one of {*first and last name*}, {*last name*, *birth month*, *sex*} or {*first name*, *birth month*, *sex*}. The second iteration identifies additional matches where the *first name, last name, birth month*, and *sex* are identical, plus either 7-8 digits of the *SSN,* or two or more of {*birth year*, *birth day*, *middle initial*, *date of death*}. It is important to note that these rules are still human provided and can easily be expressed in terms of values for the set of distance measures for different features.
When the datasets do not contain common unique identifiers, but do contain other common fields whose values can be compared (e.g. date and place of birth), matches may be found by probabilistic record linkage. Rather than assigning a discrete label 1/0 for pairs in *M* or *U*, each pair is assigned a numerical value that reflects the estimated probability that it lies in *M*.

## 2.4 Threshold-Based Probabilistic Linkage

The simplest formulation of probabilistic matching uses the frequency of different values occurring in each field in turn to compute a single match score $S_{ij}$ for two records *i* and *j* as a weighted sum of the components of the comparison vector: $S_{ij} = \sum_{k \in K} w_k \cdot D_k(ij)$.

The model codifies the human decisions in the form of thresholds as:

$$\text{Match (ij)} = \begin{array}{l} 1 \text{ if } S(ij) > S_{ij} > T_u, \\ 0 \text{ if } S_{ij} < T_l. \\ \text{'referred' if } T_u > S(ij) > T_L \end{array} \qquad (2)$$

When known matches or mismatches exist (i.e. prior probabilities), then the formulation can be changed to consider the ratio $P(D_k(ij) = z \mid \{i,j\} \in M) / P(D_k(ij) = z \mid \{i,j\} \in U)$ rather than the raw scores $D_k(ij)$ and this can easily be reformulated as a variant of Naive Bayes Learning.

Two factors are key to obtaining accurate results (i.e. to minimising the number of false and 'missed' matches).
- The first is correct estimation of the underlying conditional probability distributions, for example recognising that the relative frequencies of different family names may vary spatially (some areas may be more multicultural than others) and temporally (some names come in and out of fashion). Most approaches use variants of the Expectation Maximisation (EM) procedure to estimate the model parameters $w_k$ automatically.
- The second factor affecting accuracy is setting the thresholds ($T_u, T_L$).
- An excellent recent review of the types of errors that may be made by such procedures, and how to estimate them, can be found in Harron et al, 2017 [3]. A recent overview of the mathematical formulation of record linkage and details of the progress that has been made over the years in providing data producers with the tools to perform imputation and record linkage in feasible timescales can be found in Winker (2018) [4].

## 2.5 Machine Learning-Based Record Linkage

Simple probabilistic record linkage is limited in accuracy by the assumptions it makes, such as independence of records. More particularly, it is extremely limited by the fact that it codifies human decisions (and prejudices), so is limited by the range of possibilities a person has the time or imagination to consider.
Machine learning (ML) is a branch of Artificial Intelligence (AI) concerned with automatically inducing predictive models from datasets. Generally, given a 'training set' of observations (inputs) and corresponding labels or values (outputs), the idea is to learn a predictive model that, when presented with one of the inputs, will correctly map the input onto the corresponding output. If the training set is representative of the underlying data, the aim is that the model will also generalise, so it will predict the correct labels for as-yet unseen inputs.

If labels are not available, 'unsupervised' ML algorithms can still be used to 'cluster' the data into useful groups. Various measures of information may be optimised to seek models (transformations) under which the distance between points in the same cluster is minimised, while the distance between clusters is maximised.
While ML algorithms all incorporate certain biases in the type of models they can learn, they are generally 'assumption-free' and can exploit the power of modern computing technology to explore huge numbers of potential models in a way human designers cannot.

Recently, sophisticated ML algorithms have been applied to learn a function that projects from the space of 'comparison' vectors $S$ onto a binary (M or U) decision. In some cases, these are extended (as per the thresholding approach above) to consider a third 'manual' possibility for the output: *ask the expert*. In all cases the goal is to learn a set of decision boundaries over the space $S$, and the approaches differ as to whether any initially labelled pairs exist.

- Examples of unlinked records can usually be easily generated via blocking techniques.
- When known examples of linked records exist then these can be collated to form a training set, and provided this is sufficiently large, **supervised learning** algorithms can be applied. Many authors have reported successes learning different types of models such as decision trees (using algorithms such as C4.5 [5], or J48 [6]), instance Based Learners, or support vector classifiers [7, 8]. These are well known ML algorithms with implementations freely available in languages such as Java (via the WEKA toolkit [6]), Python (via scikit-learn), Matlab and R. The success of supervised learning approaches largely depends on generating enough linked records to provide a representative approximation of the decision boundaries in $S$.
- When no training data exists, the approach taken is to apply **unsupervised learning** in the form of clustering algorithms such as K-Means which look for natural groupings of comparison vectors. Any known examples of matches or mismatches can then provide a basis for applying labels to clusters.
- **Semi-supervised learning** is used when only a small number of matched pairs are available. This takes two forms, both of which are iterative processes.
  - If the system exploits analysts' decisions for the third (*ask the expert*) class, then this transforms the task into **active learning** [4]. This is an area of high research interest within the ML community because of its relevance to commercially valuable activities such as automated image or video labelling. The key trick of active learning is to apply a model building algorithm that provides an estimate of the model's uncertainty for unlabelled points. Learning proceeds by iteratively selecting the (set of) point(s) which are estimated to be the most informative, and then rebuilding the model after the users have provided labels for those points. An alternative approach is **self-learning**, whereby a model is trained using the initial training set, and then predictions are made for the remaining unlabelled pairs. The pairs for which the model's confidence in its prediction are high are added, to create a new larger training set, and then the process is repeated.
- Key to the success of both these semi-supervised approaches is the selection of a good sets of 'seeds' for the initial training set. The desired number of seeds can be found either by thresholding (selecting vectors with $|S_{ij}| > 1-\varepsilon$ or $|S_{ij}| < \varepsilon$, for some small $\varepsilon$ ) or by distance (sorting the comparison vectors by Euclidean or Manhattan distance from the 'perfect match' vector $\{1\}^K$ and perfect mismatch vector $\{0\}^K$) [8].

All of these three different approaches to learning record linkage models have in common the fact that the results they obtain depend entirely on the choice of the set of distance metrics ($D_1, D_2, \ldots D_K$). These define the dissimilarity between any two records, and hence the topology of the 'space' of *S* – in other words, which pairs are near each other. These choices of decision metrics can have dramatic effects on the number and complexity of decision boundaries required, and hence on the 'learnability' of the labelling problem.

- This 'metric selection' problem can be recast *in extremis* by calculating all of the possibilities for each $D_k$ and then applying well known meta-heuristic feature selection approaches. In practice it is more common to only consider a subset of the possibilities and apply an exhaustive search method using some appropriate proxy measure of the 'quality' of a set of metrics such as 'pseudo-F'.
- Common practice in machine learning is to exploit 'ensemble' learning. Also known as 'mixture of experts', this is akin to the human phenomenon of 'crowd wisdom' – the phenomenon by which a group of 'weak learners' can be combined to provide a far more accurate whole. From a machine learning perspective, the decisions of a collection of relative inaccurate 'base classifiers' can be combined (for example as a weighted vote) to create a more accurate 'ensemble classifier' provided that their errors are not correlated i.e. they make different types of errors. If 'base classifiers' are trained with the same algorithm, but use different combinations of metrics to calculate the comparison vectors (therefore learn in different versions of *S*) then this can naturally provide a diverse ensemble [9, 10].

A good recent discussion of the issuers involved in ML approaches to record linkage proposes framework that blends a variety of advances in blocking, seed selection, and self-learning ensembles, which outperforms other unsupervised learning approaches [11]. Crucially, they do not outperform fully supervised approaches (with F scores[2] around 0.95 on three of the four test sets used), although they do claim comparable performance (with F scores around 0.9 on the same three sets). These results confirm two basic points:
- (i) Provided basic cautions (against overfitting etc.) are taken, the more information provided to a system during its model-building phase, the more accurate the final model.
- (ii) The quality of matching attainable is rather high.

## 3. Attack mechanisms

### 3.1 Prevalence of attack tools

The discussion in the previous section considers existing approaches to detect record linkages used by data producers and researchers, and of course the same methods can be used by attackers. Moreover, algorithmic developments in closely related areas are progressing extremely rapidly due to their high commercial relevance. To give two recent examples: a recent summary of different ways in which semi-supervised learning is being applied to the closely related problems of identifying individuals on different social network sites is given [12], and, several recent approaches to de-anonymisation of user records combining trajectories (from GPS data) and social network location services have been outlined [13].

There is a growing movement within academic publishing for not just papers, but also related code and algorithms to be made openly accessible. Many individuals (not just academics and students) regularly get involved with competitions on sites such as Kaggle[3], which regularly include 'data cleaning' 'deduplication' and 'merging multiple data sources' as part of their challenges. As a result, it should be assumed that malicious individuals can have ready access to state of the art machine learning algorithms and process flows.

### 3.2 Inference of missing values in linked magnitude data, or data published periodically

Having established the potential for de-anonymisation via probabilistic record linkage, we now turn our attention to the threats this can pose even when some parts of records may be missing. Equally, this can be thought of as methods for linking query results across the time-series data -that is to say, the same datasets released at different times (annually etc.).

There are several different ways of thinking about linked data that lead to different mathematical formulations or machine learning approaches to attacking them. For reasons of simplicity, and also bearing in mind the major driver behind much current ML research, it is easiest to think of a simple analogy based on video processing[4].

---

[2] The F score is a measure of accuracy proportional to (1 – number of false matches)) (1- number of missed matches)
[3] www.Kaggle.com

- A video may be thought of as a time-series of images:
    - Each image consists of a number of items (pixels), each described by a number of numerical values (channels) – for example Red, Blue and Green.
    - The pixels are usually arranged or indexed in a consistent way that can be decoded to provide some spatial relationships.

- Data producers periodically publish releases of datasets. Therefore in this analogy:
    - The images might represent different time-stamped datasets.
    - Pixels correspond to cells that might have both geographical and other linkages that can be exploited.
    - Channels represent variables or sub-totals of variables within the same datasets.

A typical image/video processing problem is that a pixel can have a missing or noisy value for one or more channels, and you want to reconstruct what the actual values should be using information from the same image or the sequence. The equivalent problem in disclosure control is to remove protection by exploiting information from the same table, linked tables, or different time-stamped releases. If you have results from queries or marginal totals, then you can pose this as an optimisation problem and tackle it using mathematical programming approaches (which are exact but don't scale or deal with discontinuities) or metaheuristic search methods.

As a **spatial** problem, reconstructing a pixel's values can be tackled by considering it in the context of other pixels in the same image. If you don't have queries or marginal totals then you can solve it by trying to look at related images and learning something about the types of spatial relationships that are typically present.

- For instance, you might learn from your training set that it is incredibly rare to have a single pixel channel with a value (absolutely, or in proportion to its other channels) which is very different from its neighbours.
- Or, you might learn to recognise simple patterns such as lines or curves, or more complex textures.
- Then you can apply the model you have just learned to make a prediction of the value of the pixel of interest.
    - This is usually done by a process of 'convolution' i.e. moving small (usually rectangular) kernels across an image, where the contents of the 'kernels' define the transformation. Until 10-15 years ago the choice of kernels was the preserve of image processing experts, but this has been

---

[4] There is another reason for drawing attention to the relationship between image analysis and record linkage-based attacks: most readers will be familiar with the automated image-record linkage that happens at e-passport gates when a captured image of someone entering the country is linked to the record in a passport.

- dramatically changed by the advent of effective methods for training so-called deep convolutional networks.
  - Originally focussed on image processing, these techniques have in the last ten years shown increasing success in combining multiple data sources and making useful predictions in areas such as earth sciences. For example, if you have an industrial plant that causes air pollution, then the techniques could be applied to learn a model that linked plant output to open source data such as measurements of air quality, weather and local transport flows.
- In a data producer context, this process of applying models based on spatial relationships is how lots of 'traditional' imputation methods work, but they tend to encode certain human assumptions. With the advent of affordable scalable machine learning algorithms that can handle massive datasets and learn without assumptions, this approach could link arbitrary open source or private datasets with official statistics.

As a **temporal** problem, pixel reconstruction equates to predicting the value of its channels assuming you have previous measurements. There are well established mathematical approaches (the SARFIMA family) that can take into account long term and seasonal trends.

- More recently approaches such as (deep) recurrent neural networks are transforming the world of time-series prediction in domains such as speech recognition, machine translation etc., often exploiting 'transfer learning' (training predictors on related, possibly synthetic, datasets) to get around problems such a lack of data.
- In a data producer context, this could relate to linking publishing tables from different dates, each of which may have been fully protected, but where the protection was calculated without taking into account what was done in previous periods. In some cases, there might not be enough data to learn from, but even then, if data producers publish data without being mindful of what they did in previous years (and generally this is only done manually) then it would be trivial to write programs (that barely even count as AI) to spot unusual, missing values.

As a **spatio-temporal problem** this means learning predictive models that take into account both spatial relationships (value of other pixels or other records in tables), and temporal values (value of this and other pixels or records over time).

- Because of the interest in technologies for labelling and understanding videos, (and indeed within the entertainment industry for creating content algorithmically) this is a rapidly developing field.
- From a data producer perspective this could mean creating models using datasets that are linked over time, and across space (different sources). The datasets need not necessarily be published with the same frequency, or at the same kind of spatial resolution. There is a trade-off here. On one hand clearly there are much richer sources of data that could lead to far more informative or disclosive predictions being made. On the other hand, the space of possible models through which the ML algorithm must search becomes exponentially larger, making it harder to identify good models.

**Clearly the ability of modern machine learning to build accurate predictive models by linking diverse datasets poses a threat to confidentiality of datasets protected in isolation**. Understanding the degree of risk is a tricky issue that merits further investigation.

By their nature, these approaches are not exact, although often it will be possible to get a measure of the system's confidence in its prediction. Performance guarantees can only be estimated via their performance on data to which they have already had access – a sort of post-hoc definition. Despite the ever-increasing availability of cheap computing resources, the success of this type of approach is still likely to rely on a certain amount of expertise in identifying an appropriate formulation and representation for the problem.

## 4. Data quality and the potential for re-identification through linking

There is a difference between linking for statistical purposes, and linking data to breach confidentiality. In the former case (covered in Section 2), there is a large academic literature focusing on the problem of maximising the size and power of the linked dataset. In contrast, the equally large statistical confidentiality literature (introduced in Section 3) focuses on the feasibility of any match. In this section we consider the reality of data linkage.

In practice, linking datasets is not a straightforward issue. There are three areas to consider: how certain are we that the individuals are in the data? How good is the link field? And how do we parameterise the linking algorithm?

### 4.1 Are the individuals in the data?

For the survey data traditionally collected by the Office for National Statistics (ONS), the presence of an individual in a specific dataset is open to considerable doubt as only broad sampling mechanisms are described in the methodological bulletins. Unless an individual were to self-identify as a Labour Force Survey (LFS) respondent, for example, there would be no compelling reason to suspect that a particular person is included in the data. While there are some exceptions (for example, the exact sampling criteria for the Annual Survey of Hours and Earnings (ASHE) have been circulated in the past), there is considerable uncertainty about the composition of surveys.

This even extends to business data: although large firms are always included in ONS data, how a firm chooses to structure itself for statistical reporting purposes is not revealed. There are well-known cases (such as Rolls-Royce) where the public understanding of what a company 'does' may be different from both how the company sees itself and how ONS records it. An informal review of the identifiability of large research organisations in the Business Enterprise Research and Development (BERD) survey in the 2000s by us (unpublished) concluded that probability of a data user correctly 'guessing' the respondent was considerably lower than the probability of an incorrect guess, even for the largest companies.

The use of administrative data changes this because the inclusion criteria are known. Every individual who has claimed benefits or paid employment taxes should be in the Department for Work and Pensions (DWP) or Her Majesty's Revenue and Customs (HMRC) administrative data. Furthermore, while an individual driver may not be in the records of any specific motor insurance company, as insurance data are amalgamated from multiple organisations, the probability of inclusion approaches certainty.

Administrative data are also more likely to include individuals who do not have a current relationship with the organisation. Both government departments and commercial organisations retain information on clients after the initial transaction has taken place, on the reasonable assumption that further transactions will take place in the future.

For administrative data, then, the initial linking condition (does an individual exist in both datasets?) is much more straightforward. We would also expect administrative data to be less sensitive to the point in time that the link was made.

### 4.2 Accuracy of link fields

Match quality is highest when a unique link field already exists, but perceptions about that link field may not be accurate. For example, National Insurance numbers (NINos) are usually assumed to be unique across time and datasets. However, temporary NINos have been issued to multiple individuals in the past, although not at the same time. The impact can be seen when linking ASHE data across time, as there are individuals who repeatedly change sex and whose age varies substantially from year to year. Similarly, although reporting unit references (RURefs) for business data are supposed to be unique and one-use only, our detailed analysis of the microdata has raised some question over this.

These are exceptional examples. NINos and RURefs are generally high-quality link fields, and problems with them are easily detected. Moreover, both NINos and RURefs are of little value to attackers: NINos are replaced in ONS datasets by random identifiers as soon as linking is complete as they have no statistical value; RURefs are internal to ONS and have no external representation.

In practice, linking is most likely to be carried out using 'everyday' identifiers, where multiple values are common. Errors can sometimes be noticeable (for example, death dates often demonstrate statistical artefacts) but the appropriate correction is unknown because of the range of feasible values. On the assumption that the individuals to be linked in both datasets exist, then the likelihood of a match is inversely related to size of the dataset: a larger source dataset will produce more duplicates.

Unstructured link fields (name, address) present more problems for linking than structured fields, such as date of birth or gender, which have a limited range of values. As noted above, synonyms ("Pete", "Peter"), homophones ("Ricci", "Ritchie") and spelling differences ("Steven", "Stephen") are easily handled by humans but can still present substantial problems for rules-based match systems, even allowing for improvements in identifying variants. This is because blocking needs to be carried out before matching, which limits the opportunity to correct errors by observing the full context; for this reason, the National Health Service (NHS) recommends avoiding linking on pseudonymised links [14]. Pre-processing of link fields prior to blocking is a key challenge for linking systems, but is an area where rapid advances should be expected to produce high quality automated probabilistic matching given the volume of current research.

Administrative data are often described as 'high-quality' as it is needed for operational use, but it may be much lower in reality. Administrative data are cleaned for operational purposes, not for statistical ones. For example, the unique and high-quality NHS number is used for planned activities (treatment centre visits etc), but hospitals still use name and date of birth in interactions with patients, particularly in ad hoc situations such as accident and emergency (A&E). Harron et al. (2017) for example augmented NHS records with 'human' variables [15].

It is not easy to generalise about the quality of link fields as it is very context dependent. For example, the ONS Inter-Departmental Business Register (IDBR) has an extensive set of pre-processing rules developed over many years to enable linking of data from Companies House and HMRC; however, it is not clear that this experience would be adaptable to other environments.

### 4.3 Quality of matches

Linking algorithms will generate Type I (false positive: incorrectly linking two different individuals) and Type II (false negative: failing to link an individual who appears in both sources) errors. As noted above, the aim of the linking algorithm is to separate the combined data into 'linked' and 'unlinked' sets in a way that minimises these errors. The difficulty is that these are, by construction, largely unmeasurable: if it were known which links or non-links were erroneous, then one would expect the algorithm to correct for them.

Papers studying linking methods often test them by creating datasets with known attributes and then removing data, so that the algorithms under review can be compared against the 'true' result. Unfortunately, it is not easy to draw general conclusions or specific lessons from this, as the performance of the algorithm is always sensitive to the source data. As a result, the weight placed on these two errors reflects the goals of the linker. Both types of error can affect the inferences drawn from the paper by biasing the linked sample, but they might also have other consequences [16].

In summary, the understanding of Type I or Type II errors is similar to that of measurement error in statistical analysis: we assume that it exists, but the size and impact is largely unknowable without triangulating evidence. We also note that even deterministic linking is subject to these errors, unless the link field values are unique to the data subjects and 100% accurate.

## 4.4 Perceptions of quality

Finally, we consider what we mean by 'quality of link'. When datasets are linked for statistical purposes, a typical measure of quality is some function of the proportion of linked records to the number of expected links. For example, suppose data from a General Practitioner (GP) practice were linked to health data from the local hospital with 25% of the GP's patients being found in the Hospital Episode Statistics (HES). If this were all of the practice's patients, this might be considered a high proportion out of the possible links. If however, we were only linking patients that had taken up referrals, the proportion successfully linked is relatively low.

Unfortunately, when considering linkage for disclosure purposes these considerations are not relevant. A disclosure occurs when a record can be linked to an external dataset and used to re-identify an observation, but a high re-identification rate is not necessary for disclosure. The number of linked observations only matters in that a high number of links would indicate a systematic failure of protection, whereas a very small number of links might be the result of extremely unfortunate circumstances. In an extreme case, a single easily linked observation may be sufficient: for example, Church of England staff records containing gender records would identify the Church's first female bishop.

For disclosure, false positives are also much more problematic. A large number of linkages, whether accurate or not, generates the impression that the data are not well protected. A Department of Energy and Climate Change (DECC) study in 2014 [17] (DECC is now known as the Department for Business, Energy and Industrial Strategy) carried out an evaluation of energy data protection by asking a mix of data experts and the general public to re-identify individuals from the protected dataset, under controlled conditions. There were no successful re-identifications (DECC had access to the identified source data) but a number of false positive identifications were made. As a result, DECC chose not to release the dataset, arguing that the *perception* of re-identifiability posed a similar reputational risk as genuinely risky data.

The importance of perceptions means that the actual link quality matters much less. Recent work on public perceptions in the UK, by ADRC Scotland in particular [18], demonstrates widespread confusion in the public mind on identification risk. In general, non-specialists have a very poor understanding of what makes data identifiable; when asked hypothetical questions, the public tend to believe that data are much more identifiable than it is; but when shown de-identified data (for example, an Excel spreadsheet of numbers) the public is likely to underestimate the re-identification potential.

One way to reconcile this is to consider that members of the general public are not experts in data linkage, and so rely upon the 'expert' opinion. However, experts tend to focus on the *possibility* of re-identification, rather than the *probability* of it occurring and the associated requirement for expertise, time and resources. The difference between these two may not be evident to non-specialists. As a result, while data experts are largely concerned with the poor quality of links, the public perception may be that data linking is relatively easy.

## 5. Confidentiality issues

### 5.1 Why is it a problem?

Concerns about linked data arise from three sources: the unpredictability of future linkage, the protection of the source data, and the 'intruder' problem. The first is a problem for all data protection, but the latter two are particular problems for linked data gathered from multiple organisations.

### 5.1.1 Unpredictability of future scenarios

Before the advent of the digital economy, data management was relatively straightforward. Sensitive data could be held in protected environments. The sources of corroborating information available to a potential intruder were relatively knowable and likewise protected where sensitive; unknowable sources (such as private personal information which coincidentally matches target observations) were thought to be rare enough to be manageable.

There were still disclosure concerns. In a landmark paper, Latanya Sweeney (2000) calculated that 87% of the US Census records, and by implication the population, were uniquely identified by date of birth, gender and zip code [19]. Given the size of UK post codes, this may be even higher in the UK, and a further study in the UK used these variables plus NHS number to construct their mother-child linked dataset [15]. In the US, even at a higher county level (upwards of 15,000 individuals), it was estimated that 18% of the US population were unique on date of birth and gender, and Sweeney has spent two decades demonstrating how commercial data holders, in particular, routinely share information at this level or more detailed. Nevertheless, the argument that data sharing could be predicted and controlled to a reasonable degree was a fair assumption.

However, the development of the internet over this century has demonstrated how difficult it is to predict new digital developments, particularly in respect of social media and mobile phone data. Facebook, Twitter, and WhatsApp collect both static data (names, date of birth) and dynamic data (events attended, meals, meetings etc); the latter are less identifiable in themselves but can provide highly accurate matches because of the uniqueness of personal activities. The popularity of using Facebook to log in to other services greatly increases the dependency of individual privacy on a single point of failure. Operational data collected through mobile phones and made available to apps can be extremely revealing: it has been shown that as few as four location timestamps are enough to identify 95% of individuals in a dataset of 15 million, even when the time/date resolution is relatively coarse [20].

Of most relevant to this article, however, is the unpredictability of these services. At the end of the last century, the expectation was that smarter use of cookies and scripts, along with server-side web page delivery such as Active Server Pages, would drive digital technology into ever-better customer focused services. Business texts of the period very largely take the existing non-digital business models and add technology [21, 22]. An implicit assumption of the period is that companies would jealously guard information as this would give them a commercial advantage. Social media as we know it today, and the implications for data privacy, was completely unanticipated.

The theoretical problem of protecting data against unknown future scenarios is obvious. One practical problem that receives less discussion is the recall problem: how do you deal with recall of datasets or statistics that were deemed safe under an old regime when the world changes? For example, suppose the UK followed the Norwegian example of making individuals' salary and tax details available online. This is probably sufficient to make observations in the LFS End User Licence files identifiable with minimal effort. These files are currently widely distributed amongst academics; would the ONS be under an obligation to try to remove all those files from circulation? If so, how could the ONS be sure it has done so? Just asking users to destroy files might encourage others to try to find out what was so sensitive in them.

This is a potential problem for all microdata and statistical releases. In the context of linked datasets, an additional problem is that data holders need to consider not just potential risks to the data as a whole, but to each contributing part, which takes us to the second problem.

### 5.1.2 Protection of the source data

By construction, linked datasets are generated from two or more separate data sources. This creates scope for different levels of protection of the source data. This could arise from differing levels of security assigned to source data by the same organisation (for example, at the ONS, producer price microdata are confidential while retail price microdata are not), or it may be because different organisations apply different data standards (for example, commercial firms are likely to rely upon contract terms to protect data whereas the ONS relies on operational measures). A risk then arises because the less protected source data might allow more sensitive linked data to be re-identified.

Consider data being held at the level of primary care trust (PCT) and at the individual level. The PCT has to produce a set of outcome measures for public consumption, but the personal data are deemed confidential and de-identified for secondary users. Linking the two may then inadvertently increase the likelihood of re-identification.

Similarly, the use of company accounts to supplement confidential ONS data leads to a public and identifiable information source being linked to a confidential, supposedly non-identifiable one. The cardinal numbers in accounting information are very likely to be unique to the organisation. As the industrial sector of the company is almost always included with business data (without it, the data are of little value), even approximate or rounded numbers may be unique.

All government bodies (and some non-government bodies) should apply a common security standard known as CESG (Communications-Electronic Security Group, absorbed in 2016 into the National Cyber Security Centre). This largely relates to IT equipment and procedures, but there are multiple domains, 'Five Safes', to data sharing, including the statistical domain [23]. The Five Safes model is used internally by the ONS, and is reflected in the Digital Economy Act 2017 (DEA). The DEA allows for certification mechanisms for the non-statistical domains (purpose, users, settings) for all government data with the exception of health. In theory this provides a straightforward way for government departments, at least, to develop shared access regimes and common standards. However, although the ONS has begun actively promoting the Five Safes and DEA in this spirit, it is too early to say how effective this will be.

### 5.1.3 The intruder problem

SDC research uses the 'intruder' as its basic risk scenario. This is an individual who has malicious intent to breach data security, technical knowledge, sufficient resources, and crucially for our purposes, access to corroborating information. Protection against the intruder is the cornerstone of SDC.

This model has been strongly criticised and it is argued that this approach fails to appreciate genuine risks in data access, over-compensating for hypothetical risks [24]. In particular, they criticise the standard assumption about what external data an intruder might use to break data protection. Papers which attempt to match units across different data sources typically find that matches are very low quality, even when the same conceptual unit is being discussed [25, 26, 27].

However, in the case of linked datasets acquired from different organisations, this criticism does not hold. By construction, at least one organisation other than the linked data holder has access to a **complete and exact** copy of the data used to create the linked dataset. This is a far riskier situation than that usually posited in SDC models. Moreover, if the source data are derived from administrative records, the intruder is likely to have an accurate idea of which individuals can be reasonably found in the dataset, and also likely to be able to link direct identifiers to that data.

### 5.2 Threats to confidentiality arising from existing practice in publishing magnitude data

Statistical summary data can be extremely vulnerable to attack now that computing power and algorithms exist to link and attack large datasets e.g. same datasets released at different times (time-series data).

Threats to the confidentiality of published magnitude data arise from two sources. The first of these is inadequacies in widely used tools. Existing methods for cell suppression rely on either exact mathematical programming algorithms, which do not scale for tables larger than a (very) few thousands of cells, or on heuristics such as HyperCube which can be shown to simultaneously over- and under-protect [28]. As a result, data producers manually 'partition' larger tables into more tractable sub-tables, which they then protect and recombine. The risk here is that patterns of linkage across sub-tables can lead to confidentiality disclosure risks.

As an example of the risk involved in current practice, even when everything is under the data producer's control, in 2011 the University of the West of England (UWE) undertook an exercise to 'attack' a set of well-known statistics that had been published as a series of 2-dimensional tables protected by cell suppression using algorithms within the Tau-Argus software. Using only the published data, the tables were first recombined into one large table. Then an 'unpicker' tool authored by UWE was applied which iteratively tightens upper and lower estimates of missing cells' values. **The results were that all of the missing cells could have their values calculated to the same precision as the rest of the published dataset.**

This failure of existing practice was reported to the data publishers and the authors of Tau-Argus. Investigation revealed that it arose from a combination of manual partitioning and the use of inadequate heuristics within Tau-Argus. We have subsequently repeated this exercise with synthetic data and published the findings to warn the community, and the authors of Tau-Argus [28]. The fact that we were using data identifiers from the data publisher, and there was a consistent methodology, meant that we could link records with certainty. However, it would be relatively simple to allow for quantified uncertainty, and the results would be slightly looser bounds in some, but not necessarily all, cells.

However, Artificial Intelligence can offer solutions to this problem. Smith et al. describe the development and benchmarking of an approach which combines 'evolutionary algorithms' with mathematical programming and can **safely** protect tables orders of magnitude larger than other algorithms [29]. Tools implementing these algorithms are now being used to protect (and validate the protection of) some business and trade statistics at the ONS. Advances in the field of 'hypergraph' partitioning offer the prospect of safer, automated break-down of huge (possibly linked) datasets which would greatly extend the applicability of approaches such as cell suppression algorithms.

There is quite a substantial amount of protection in aggregate statistics (which includes both frequency and magnitude tables), however. The records themselves inherently contain uncertainty due to sampling methodologies (although as noted above this may not apply to administrative data) and the fact they are a snapshot in time. For example, ONS statistics are generally weighted population estimates (except for the Census and IDBR statistics) rather than sample descriptions; the weighting introduces another level of uncertainty, particularly if those weights have been changed post-sample to allow for quality problems. Finally, aggregate statistics typically come with low dimensionality, reducing the potential for identification even when multiple tables are linked.

### 5.3 Threats to privacy from microdata releases

For microdata, the intruder's problem is much simpler. The microdata are personal and multidimensional, and any unique values or combinations of variables can be quickly identified. The range of potential link fields is very large. For example, suppose a friend states that she has completed the Labour Force Survey in 2017. On the reasonable assumption that a friend would know her age, occupation, level of education, marital status, and employer's industry, it is likely that it would be easy to identify her in the dataset. If she was known to be from a relatively rare ethnic background or nationality, the link can be achieved with much less information. Finally, if the data are created from linked data from multiple organisations, then the problems noted in Section 5.1 exist. An intruder with access to the microdata from the supplying organisation, whether acquired legitimately or not, should be able to directly re-identify records.

The anonymisation techniques described in other articles and the wider literature can reduce the likelihood of re-identification. The difficulty is that the anonymised data must retain some usefulness: *"the richer the dataset, the greater the set of possibilities for useful auxiliary information, and a host of results suggest that de-identified data are either not de-identified or no longer can serve as data"* [30]. Some personal data can be usefully de-identified, but business microdata are assumed to be only unusable or identifiable with no middle ground, as the key identifying characteristics (size and industry) are the variables of most value.

Moreover, time is against the data holder: "*Anonymization of a data record might seem easy to implement. Unfortunately, it is increasingly easy to defeat anonymization by the very techniques that are being developed for many legitimate applications of big data*" [31]. Time makes re-identification increasingly likely whatever the sophistication of statistical methods.

**Figure 1: Identifcation risk and technological advances over time**

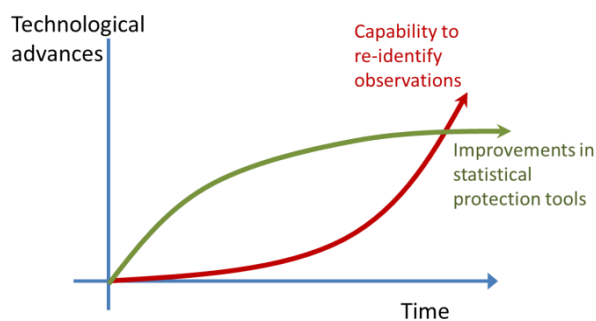

Data availability, processing power and the efficiency of re-identification algorithms have been increasing exponentially, and there seems all likelihood that this will continue; if nothing else, brute force algorithms become more feasible over time. In contrast, anonymisation techniques are likely to see a decline in the effectiveness of new techniques as anonymisation needs to be balanced against the impact of protection measures on usability; usefulness places an upper boundary on how far data can be protected.

Unknown future technological developments may affect both, but it seems more likely that re-identification techniques will continue to improve at a faster rate as these are unbounded. In short, the historical perspective argues that at some point, anonymisation, while retaining useful information, must become infeasible. Linking data from multiple sources is likely to hasten that end point.

## 6. Methods for preserving privacy in microdata release

Section 5.3 suggested no future for microdata release: how can you prevent re-identification in a world of unlimited challenges? But this is taking a data-centric view: we have some data; how can we best protect it? Best-practice is moving away from this to a user-centric view: what information needs are we trying to meet with our data? This focuses the mind on the first part of the usefulness-anonymisation trade-off, rather than the latter. In addition, it allows one to consider non-statistical solutions to data problems.

For example, consider the increasing public demand to be able to create customised tables from the Census. This user need could be addressed by:
- Offering a bespoke table-creation service
- Creating an automatic tool to generate tables
- Releasing an anonymised version of the Census microdata
- Releasing synthetic microdata which produces approximately the same tabular results

Each of these has pros and cons; therefore, user input is crucial into making the correct investment. The value of this is recognised in recent legislation (such as the Digital Economy Act 2017 and the General Data Protection Regulation (GDPR)) which recognises that multiple legal, technical and procedural measures can be combined with statistical processes to provide safe data outcomes.

In this section we consider how both statistical and non-statistical solutions can help to meet user information needs whilst still protecting data.

### 6.1 Statistical solutions

#### 6.1.1 Traditional anonymisation techniques
Despite the comments made above, traditional anonymisation techniques have a long history of usefully balancing risk and utility, and they continue to work for many current environments. Analyses such as those discussed above focus on the worst cases, assuming an attack. However, even if we allow for the possibility of an attack taking place, this does not mean that it will happen, or is even likely to.

A key assumption of worst-case modelling is that an intruder has both skills and incentive to re-identify data. Consider an intruder who works for an organisation that has supplied data to the data holder. If the intruder could see the unprotected microdata, he should be able to re-identify the individuals. However, relatively simple measures to swap records, remove outliers, round or limit values may be enough to confuse all but the most experienced data analyst.

This does not address the problem of future re-identification in a dataset which was released under the belief that it was safe. However, the passage of time does provide some protection: the identifying information may have changed, and the information itself may have lost value over time.

This illustrates a wider problem with the 'everything-is-identifiable' approach: it assumes that the value of the data is sufficient to justify the time and resource cost to link and re-identify the data. The typical statistical disclosure control (SDC) model assumes that an intruder has a powerful motivation to re-identify data, and usually the motivation is to undermine the data provider. This motivation is useful, because it needs no further justification with reference to data value, and directly places re-identification as a contest between two parties fighting over the same goal.

In the real world, it is not clear whether this is relevant. There are clearly some cases where the knowledge gained is valuable enough to warrant extensive attempts at re-identification; abortion data are one example in the UK, and there are concerns in some other countries that official data are attacked to glean information for extortion purposes. However, it is not clear that these exceptional cases should provide a benchmark for risk for all datasets; on the contrary, public benefit is likely to be better served by treating these cases as exceptions, and considering more reasonable motivations.

Re-evaluating motivation can also uncover alternative protection mechanisms. For example, Eurostat accepted arguments [32] that the commercial value of business data contained in the Community Innovation Study (CIS) was so low it made re-identification for personal gain meaningless. Of course, re-identification to undermine Eurostat was still possible as this concerned business data with industry and size included in it. However, it was argued that, as the data was released under licence, protection was adequate [32]: an attacker demonstrating that a business could be re-identified would achieve nothing. The business was acknowledged to be identifiable, which is why it was released under licence; re-identification for commercial gain was valueless; re-identification to prove re-identification on a dataset known to be re-identifiable makes no sense. Had the data significant commercial value, then the possibility of distributed data leaking out into the wild beyond licensed users would need to have been considered, but the lack of an incentive to re-identify made the de-identification decisions much easier.

In short, while theory suggests distributed data cannot be anonymised, practical evidence suggests it can. This is recognised in the German Federal Data Protection laws, which employ the notion of "de facto anonymisation" as the protection standard. That is, a dataset is anonymised if the benefit gained from re-identifying the data is outweighed by the time and resources needed to carry out the anonymisation.

### 6.1.2 Synthetic data

Synthetic data involves the replacement of some, or all, of the original source data by values drawn from an artificially created distribution. To retain analytical integrity, multivariate distributions are used to create the draw pool. A fully synthetic dataset has all the values replaced by synthetic equivalents; in a partially synthetic dataset, only 'risky' (that is, more identifying) variables are replaced. The advantage of the synthetic data is that it allows one to circulate microdata which is safe but approximates the real data.

In Europe, much of the work in this area has been in Germany [33, 34] and in Scotland [35, 36]. These teams have developed general purpose R code (SimPop and SynthPop, respectively) to create synthetic data files automatically. Most of this work is focused in creating public-use files (PUFs); that is, files with no restrictions on release and circulation.

In the US, synthetic data files were developed to be used in online versions of secure facilities [37]. Researchers have used synthetic data to generate analyses which are suitable for publication in peer-reviewed journals. As a result, much more research has been done in the US in the validation of results [38], including online services which provide both results and confidence measures. The US has also pioneered the development of synthetic business data [39].

Synthetic data are intended, at least in Europe, to meet the data needs of individuals who require access to plausible detailed microdata but do not need the specific values in the original data; an approximation to results is sufficient. This meets three user needs:

- Allowing non-specialist users to generate their own approximate analyses
- Teaching
- Code development for those working on the sensitive data in restricted-access environments

In the US, confidence in the analytical validity is such that even specialist users are working on this data.

There is still some theoretical disclosure risk. For partially synthetic data, some of the non-synthesised variables might be identifying. In addition, the more closely the synthesis represents the original data, the more it reproduces the original disclosure risk. On the evidence to date for released synthetic data files, these risks seem relatively small. Synthetic data files also have the advantage that they are less likely to be problematic in the future: as the data are artificial, better matching in the future is not relevant.

## 6.2 Non-statistical controls

### 6.2.1 Query servers

Query servers allow users to submit a request and receive a statistical response, without ever seeing the data. Table query servers have been offered by statistical offices for many years, meeting the public demand for more flexibility on table production that the data producer is willing to provide.

In table servers, confidentiality is managed either by perturbing results, or by limiting outputs to a known finite set of variable combinations. However, with the growth in resources for an intruder, and the wider range of tables being requested, concerns have focused on the possibility of a breach using the sort of techniques described in Section 5.2. As a result, the current generation of table query servers being developed in Australia and the UK, build in noise which can be applied consistently across multiple tables, removing the chance of unpicking the noise through multiple table differencing. The downside of this technique is that tables are no longer additive, which may confuse non-technical users who may see inconsistencies between results.

A special case of the table query server is a differential privacy (DP) server. This guarantees a minimal level of uncertainty about the true source values, by adding sufficient noise to generate this uncertainty. This works well for frequency counts when both the population and samples of interest are large but can give very misleading results for rare events, or when applied to magnitude tables. In theory, repeated table requests can be used to home in on the true values: DP is only guaranteed in the context of a finite number of requests. However, as DP advocates note, this is true of any system of random perturbation to a fixed set of rules; in practice, unlimited repeat queries are easy to track and organisations block this.

Analytical servers, allowing users to run more complex, but finite, sets of analyses are rarer. Unlike remote job servers (see Section 6.2.2 below) which allow fairly unrestricted but indirect access to microdata, an analytical server restricts the user to a fixed set of commands. Careful selection of the commands can cover much of the ordinary analysis required by users. Unrestricted coding can then be limited to a very small group of very expert users.

An outstanding example is the service launched in 2018 by the Norwegian Statistical Office[5]. This allows users to register with minimal intervention using national personal identification numbers. Registered users can then write, run and save Stata-like code, create variables, save datasets (within the system; no data leaves), produce univariate and multivariate analyses, and swap code between users. As analytical use is not often in conflict with SDC principles, the Norwegian system applies simple broad rules which provide a high degree of confidence in the non-disclosiveness of outputs, while also not overly restricting research.

---

[5] www.microdata.no; currently only available in Norwegian.

### 6.2.2 Remote job servers

Remote job servers (RJSs) accept code from researchers and run it against the data, returning the output of the code. The advantage for the data holder is that researchers do not get to see the source data, and every request for data and statistical output can be recorded and checked if required. The best example of this is the LISSY server,[6] run for fifteen years by the London School of Economics to analyse the Luxembourg Income Study.

The confidentiality risk in RJSs is debatable. Some authors have argued that users are fundamentally untrustworthy, and so the data need to be highly protected [40], while others argue that user problems are manageable by having properly designed systems [41]. The evidence is ambiguous: while LISSY for example has had no confidentiality breaches in over 10,000 analytical requests, the Australian Remote Access Data Library (RADL) suffered a number of deliberate co-ordinated attacks. Note however that these attacks were not designed to re-identify data, but to give researchers more control over it.

What is clear is that RJSs do protect against mass linking of two datasets via technical measures: users have no ability to upload additional data sources. A user could try to compare individual data points to some other information such as social media data, but this is not straightforward. As the source data are not made available, the attacker has to collect this information indirectly, for example by using differences between tables. Such patterns of behaviour do not reflect genuine research use, and are likely to be identified quickly by the RJS manager. Because RJSs store all the code requests, it is straightforward to demonstrate that a user has acted deliberately and unlawfully.

### 6.2.3 Remote research data centres

A research data centre (RDC) is a facility that allows the researcher largely unlimited access to data, but under the control and supervision of the data holder. A remote or virtual RDC (rRDC) has similar functionality but allows the researcher to access the data from a location separate from the data holder. Security is achieved through using thin client architecture to retain all control over all practical functions at the data holder, with the research only having a 'window' onto the data.

Generally, security is maintained at a very high level; yet the rRDC is also popular with researchers as it gives them complete access to the data. As a result, since 2003 this has become the standard way across the world to provide access to sensitive microdata; the ONS' Secure Research Service is an example. Other examples, with varying degrees of flexibility, include the 'safe pods' installed by the Administrative Data Research Network at UK universities; the UK Data Archive's Secure Data Service which supplies ONS data to UK university desktops; the 'hardened thin clients' distributed by the French CASD research centre[7] which allow them to offer access to their servers from any world-wide location; and the web-based rRDC 'SURE' run by the Sax Institute in Australia which allows web-based access across a range of countries.

---

[6] https://www.lisdatacenter.org/data-access/lissy/

The advantage of RDCs and rRDCs is that a very precise degree of control can be exercised over users. This greatly reduces the risk of re-identification by linking data. As with remote job servers, linking of additional datasets is usually blocked by the technical systems. Re-identification is theoretically possible by comparing information displayed on a rRDC with information on another screen connected to the internet. However, genuine research has a distinctive pattern of data use which is quite different from searching to re-identify individual data points. As all rRDCs record activity, detecting and demonstrating unlawful behaviour is straightforward.

Remote RDCs are a specialist service which only meets the needs of a small part of the population; indeed, the smallness of the user group is what allows security to be tailored to this group. However, this group are also some of the most powerful users of sensitive data. For example epidemiology relies upon the existence of secure research facilities, and the growth in rRDCs through organisations such as the Administrative Data Research Network and the Farr Institute[8] in the UK, the Centre for Health Record Linkage[9] (CHeReL) in Australia, and the INDEPTH[10] and ALPHA[11] networks in Africa has led to an explosion of public health research in many countries. Linking health data in secure but flexible research facilities is generally the example used to demonstrate to the general public the benefits of combining data sources.

### 6.2.4 Output SDC and re-identification

The rRDC movement has generated a number of knock-on effects, including Active Researcher Management [42], the development of best practice researcher training (for both rRDCs and distributed data), and the 'Five Safes' framework for data access which has been widely adopted in the UK and abroad. Most importantly for this article, it is the source of the general field of output SDC (OSDC).

OSDC arose from the recognition that (1) SDC models of tabular data protection designed for the industrial production of statistical aggregates do not serve the tabular data protection needs of researchers and analysts (2) much research output, and certainly the most important, is not tabular or linear in form. The lack of clear guidance led to the ONS developing a set of rules in the 2000s, which were reviewed by an expert group commissioned by Eurostat, and subsequently adopted by that body as best practice [43]. This document forms the core of OSDC as recognised by most research organisations which have any formal guidelines.

---

[7] https://www.casd.eu/
[8] http://farrinstitute.org/
[9] http://www.cherel.org.au/
[10] http://www.indepth-network.org/
[11] http://alpha.lshtm.ac.uk/

A key development in OSDC was the introduction of the idea of 'safe statistics; that is, reviewing the mathematical characteristics of statistics to determine whether there is any meaningful disclosure risk. The benefit of this approach is that it allows facilities to classify whole groups of outputs as 'low risk', irrespective of the data used to generate them. Tabular data protection, as practised by statistical organisations, can be then be seen in context as a particular class of 'unsafe statistics' (linear aggregations) with a particular objective (temporal and cross-tabular consistency).

Output SDC has a fundamentally different epistemological model compared to traditional table and microdata SDC. It recognises that data uses are generally highly subjective, affecting the data selection, manipulation, and interpretation, and the publication of outputs (tabular outputs from statistical organisations, being produced to a tightly-defined set of rules, are an exception). Determination of input data and manipulations are extremely difficult, as the large literature on failures in replication shows. In the past we have tried reproducing research results using datasets and code supplied by researchers, and been unable to do so. This undermines a core element of the standard 'intruder' model, the predictability of the input data, and the actions of the statistics producer.

This also leaves the OSDC community with a problem: how can SDC procedures such as the Eurostat guidelines [43] be defined when everything is mutable and context-specific? The answer lies in combining evidence, theory and expert judgment in varying measures to generate broad rules with a high degree of safety plus guidance on exceptions. For example, the statistic known as a Herfindahl concentration index has, in general, no significant disclosure risk, irrespective of the data used to generate it. There is a special case where it is disclosive, but this can be checked by someone with no technical knowledge. On the other hand, the way that these indexes are used by researchers suggests there is no risk even if the test is not carried out. In contrast, a frequency table is fundamentally high-risk, but, given the subjective nature of analytical outputs, the evidence to date suggests that a relatively low minimum cell frequency seems to provide adequate protection without limiting analysis.

What is not known is how well these guidelines manage risk in practice. Traditional SDC modellers dislike the characterisation of the risks in a research environment as fundamentally unknowable, but have struggled to develop alternative, deductive, models which match modern OSDC's ability to deal with large volumes of output quickly and safely. Meanwhile, advocates of the modern OSDC approach worry that the inductive approach might be missing crucial risks.

**6.2.5 Confidentiality risk from linking outputs from multiple sources**

This discussion of different delivery mechanisms raises the question of whether outputs from multiple sources can be linked to re-identify source data. This could be particularly relevant if parts of the source data are spread across different systems. Each of the delivery mechanisms discussed in the previous subsections treats outputs differently but provides a degree of protection.

For query servers, the addition of noise is designed specifically to prevent matching against one's own data, let alone a secondary source. For the modern table servers discussed in Section 6.2.1, even using one's own information to uncover the noise associated with an output from the table server does not help. As the noise is generated independently for every statistic, all the intruder knows is how much noise has been added to his or her own data. For differentially private servers, the whole basis of the model is that the result is protected against any other possible information, including a complete copy of some of the source data being held elsewhere. It is also clear that the risk only arises from holders of the contributing data: the purpose of a table server is to meet all feasible user needs, and so there is no necessity to duplicate data sources.

For RJSs and RDCs, the issue is different. It is quite likely that the same dataset would be spread amongst multiple delivery outlets. For example, the ONS's Secure Research service and the UK Data Archive's Secure Data Service share many of the same ONS datasets. This raises three possibilities.

The first is that an attacker could deliberately create outputs to breach confidentiality (for example, by producing a table in one RDC and the same table plus one extra observation in another RDC). This can be dismissed as it is theoretically possible but makes no sense. The attacker has access to the source data; there is no gain (and considerable risk) in trying to hide disclosures in output.

The second option is that an attacker studies outputs produced from the different facilities and compares the two for disclosure by differencing. This is different to comparing statistics from the same facility produced by two different researchers, as the different facilities might have different rules (e.g. a minimum table cell size of 3 rather than the more common 10 or more). However, it is not clear what an attacker would get out of this. The useful information is all contained within the facilities with lower cell counts, and the other facility's outputs add nothing.

The third option is that the attacker holds some subset of the data (perhaps she contributed administrative data) and compares that to output from the research facility. The aim would be to see if any additional information can be gleaned from the data. Suppose the attacker contributed 1000 employment records, and the analyst links health data and produces statistics to show that all 1000 records had a heart problem; the attacker now knows that every employee she sent data on has a heart problem. If the analyst cleans the dataset, drops ten observations and produces tables to show that 990 have a heart problem, the attacker faces some uncertainty but can be reasonably sure that any one of the employees she submitted has a heart problem.

This is clearly a potential problem, but difficult to quantify. OSDC is designed to guard against these sorts of issues: for example, guidelines for researchers typically tell them to avoid full or empty cells unless these are 'structural' (i.e. contain no new information). We also noted that replication (and therefore differentiation) of analytical results is difficult and unlikely to be successful. Perhaps the most that can be said is that, as (a) we are considering statistical results not microdata, (b) modern OSDC guidelines are designed to allow for a broad margin of error, and (c) analytical use is idiosyncratic, at this time the risk seems manageable.

### 6.2.6 Non-technical controls

As well as statistical controls and environmental solutions, data protection is affected by legal constraints and contracting arrangements. This is where private sector companies typically focus when discussing data sharing, and the effectiveness of data protection is likely to be related to the commercial value of the data (and any possibility of fines by the regulator). In our experience, contracts without additional controls tend to be a relatively poor way to protect data, as they are often disassociated with the actual data management practices and do not allow for the possibility of human error. In contrast, procedural agreements tend to work much better and be responsive to error.

It is possible to improve data security by investing in user training. A survey of RDCs carried by for the Australian Department of Social Services showed that all RDCs saw some form of user training as important for maintaining security [44]. However this is again a specialist market. It is not possible to train all users of open data, and data practitioners generally assume that the recipients of confidential data files never read the guidelines on how to look after the data.

### 6.2.7 Discussion

Unlike the data-centric approach, the user-centred approach is more difficult to manage as it has multiple dimensions: what combination of setting, user and contracting arrangement will meet this particular user need? This flexibility however also allows more targeted solutions to be derived.

The multi-dimensional approach can also help to illuminate where problems lie and can be tackled. For example, both RJSs and rRDCs allow users to access sensitive microdata from a remote location. As these are used to house the most readily identifiable microdata, a possible attack scenario would involve an intruder browsing the internet at his home location while simultaneously interrogating the data remotely. This seems like a higher-probability risk than an intruder trying a brute-force attack to re-identify distributed anonymised files. One solution would be to reduce the detail in the files, but this would disadvantage all users of the remote system, whether well-behaved or not. The multi-dimensional user view correctly identifies the real problem:

the way that users have been accredited and trained to use the system. Attention can then be focused on whether the accreditation process, for example, can be tightened up.

One key element is the difference between distributed data (sending out data for users to hold and manage) and distributed access (allowing remote use of data but retain control over it), and the growing preference for the latter amongst data producers. Query servers, RJSs and RDCs are all examples of distributed access: they retain control over the data and have more or less control as to how the user interacts with it. This has one enormous advantage over distributed data. By being able to monitor all access, problems such as repeated differencing or attempts to link external data can be identified quickly. In contrast, the data holder has no idea if a dataset she has released is being linked to something else, or subjected to a brute force attack.

Distributing access rather than data also tends to protect against an uncertain future. If it turns out that the LFS data in the ONS's Secure Research Service is creating an unacceptable disclosure risk, it is easy to withdraw that dataset from use; the same cannot be said of the distributed version.

A final issue that is coming to the fore is the epistemological differences between groups of SDC experts. For example, traditional SDC analysts looking at outputs from research are concerned about disclosure by differencing: theoretically, this exists, and the number of tables produced by researchers is rising and uncontrolled. Modern OSDC advocates, in contrast, argue that the profusion of tables is not a risk: output rules are strong and cautious, and if anything rising numbers of outputs protects the data more by creating a 'sea of confusion'. Moreover, it is not clear if research in this area can resolve the fundamental differences. For example, we are currently engaged in a project to see if there is any genuine risk in the differencing of outputs, using the intruder's best-case scenario of many users of the same dataset. But whatever the outcome, this cannot solve the fundamental conflict between the theoretical perspective ("you can't prove that you're right") and the practical perspective ("you can't find evidence to show that I'm wrong").

## 7. Summary

The theory of linking datasets is well-understood; in practice it is more difficult. This article has investigated this, paying particular attention to the disclosure risks: linking datasets to uncover information is much easier than linking data to get a high-quality matched dataset.

This is particularly the case if one of the sources is taken from administrative data, as this implies that an organisation other than the final data holder has a complete copy of the contributed data. It might also be the case that the contributing organisation does not protect the data as well as the final data holder, perhaps because some for the source information is seen as less confidential. Statistical organisations may want to consider how they encourage data suppliers to have shared data governance and transfer mechanisms – the Digital Economy Act and the 'Five Safes' framework provides the legal and practical mechanism to do this.

As well as the use of administrative information, a major concern for statistics producers is the availability of other data sources which could help to re-identify the data, and the computer power and analytical tools to do so. These are not trends we would want to stop: the use of big data, machine learning, massive simulation and so on are producing major benefits in public information, planning and policymaking. However, it is unfortunate that the great gains we see in using data are exactly what a malicious intruder could use to reduce data value. Moreover, in many cases the method and toolkits needed are available in the public domain. As a result, it should be assumed that malicious individuals can have ready access to state of the art machine learning algorithms and process flows.

This does pose some risk to statistical outputs. Official statistics are likely to be targeted by linking attacks, as are the outputs from analysis and research, but the aggregate and partially subjective nature of these will always provide some protection, even in an uncertain future world. While standardised tables might be problematic, delivery systems such as table query servers and techniques such as evidence-based OSDC appear effective in staving off problems for a considerable time.

In contrast, there are big implications for the distribution of microdata where the identification potential seems likely to increase exponentially. This suggests that rather than focusing on the problems of data protection per se, we should adjust our focus to look at it from the user perspective: what information do they actually want, and what will they do to get it? This takes us into a number of interesting and more future-proof avenues including query servers (such as the Norwegian model), distributing access rather than data, and taking an empirical approach to risk management. Further research in this area would be warranted, particularly as the Government Statistical Service (GSS) already has access to leading expertise in these fields.

We have also identified some areas where existing practice in the protection of magnitude data can be improved in the short term. An over-reliance on toolsets that do not scale up reliably to the size of problems often tackled, means that a potentially catastrophic cocktail of manual intervention and flawed heuristics is often employed. This can result in the publication of statistically tables that can be trivially relinked (over time and/or geography) to expose confidential data. Artificial intelligence based approaches are known and can ameliorate this situation to some extent, but more work is needed to generalise their application beyond a few real-world case studies at the ONS.

We have also identified areas that we believe merit targeted investigation to yield medium-term gains, such as the use of 'hypergraph partitioning' to allow the protection of larger linked tables, and an investigation of the risks posed by the application of algorithms developed within the field of image and video processing. We would welcome further research in these areas as well.

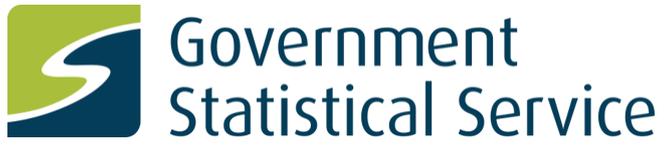 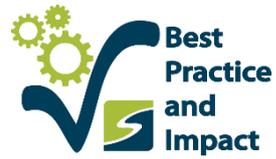 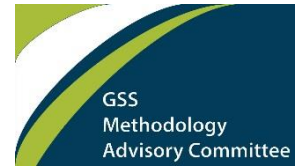